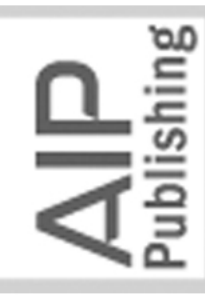

# X-ray Nano-Imaging of Defects in Thin Film Catalysts via Cluster Analysis

Aileen Luo[1, a)], Oleg Yu. Gorobtsov[1], Jocienne N. Nelson[2], Ding-Yuan Kuo[1], Tao Zhou[3,4], Ziming Shao[1], Ryan Bouck[1], Mathew J. Cherukara[3,4], Martin V. Holt[3,4], Kyle M. Shen[5,6], Darrell G. Schlom[1,6,7], Jin Suntivich[1], Andrej Singer[1]

[1]Department of Materials Science and Engineering, Cornell University, Ithaca, New York, 14853, USA

[2]Laboratory of Atomic and Solid State Physics, Department of Physics, Cornell University, Ithaca, New York, 14853, USA

[3]Advanced Photon Source, Argonne National Laboratory, Lemont, Illinois, 60439, USA

[4]Center for Nanoscale Materials, Argonne National Laboratory, Lemont, Illinois, 60439, USA

[5]Department of Applied and Engineering Physics, Cornell University, Ithaca, New York, 14853, USA

[6]Kavli Institute at Cornell for Nanoscale Science, Ithaca, New York, 14853, USA

[7]Leibniz-Institut für Kristallzüchtung, Max-Born-Str. 2, 12489 Berlin, Germany

## Abstract

Functional properties of transition-metal oxides strongly depend on crystallographic defects; crystallographic lattice deviations can affect ionic diffusion and adsorbate binding energies. Scanning x-ray nanodiffraction enables imaging of local structural distortions across an extended spatial region of thin samples. Yet, localized lattice distortions remain challenging to detect and localize using nanodiffraction, due to their weak diffuse scattering. Here, we apply an unsupervised machine learning clustering algorithm to isolate the low-intensity diffuse scattering in as-grown and alkaline-treated thin epitaxially strained $SrIrO_3$ films. We pinpoint the defect locations, find additional strain variation in the morphology of electrochemically cycled $SrIrO_3$, and interpret the defect type by analyzing the diffraction profile through clustering. Our findings demonstrate the use of a machine learning clustering algorithm for identifying and characterizing hard-to-find crystallographic defects in thin films of electrocatalysts and highlight the potential to study electrochemical reactions at defect sites in operando experiments.

[a)] Author to whom correspondence should be addressed. Electronic mail: al2493@cornell.edu

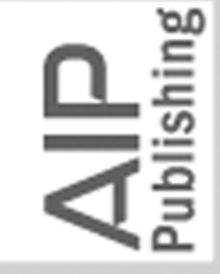

Limited natural resources and increasing demand for sustainable energy create a need for efficient electrochemical energy conversion and storage devices, such as batteries, fuel cells, and electrolyzers. Crystallographic defects are critical for a broad range of material functionalities[1], and the role of defects in electrochemical systems has attracted considerable attention. For example, the crystallographic orientation of electrocatalysts significantly impacts catalytic activity due to the differences in electronic structures associated with surface termination facets[2,3,4]. Additionally, a manipulation of the surface concentration of the A-site cation in $ABO_3$ perovskites during synthesis mitigates activity inhibition due to point defects in the form of surface cation segregation[5].

Unlike facets and homogeneously distributed vacancies, the effect of crystalline line defects (dislocations), partial amorphization, non-uniform film thickness, or point defect clusters on catalytic activity is poorly understood. In non-oxide materials, the strain gradients around dislocations were discussed early on[6]; however, the lack of operando access to structural information has prevented further forays in this direction, partly due to challenges in developing an operando imaging method with a resolution of tens of nanometers to probe the distortions produced by dislocations[7]. While environmental and operando transmission electron microscopy (TEM) provide atomic resolution information on various catalytic materials[8,9,10], side reactions with the electron beam in liquid cells may affect the systems' dynamics, and typically, only small specimens can be investigated. Surface-sensitive operando atomic force microscopy (AFM) measures changes in surface topography and electrochemical potentials during energy conversion processes such as the oxygen evolution reaction (OER)[11,12]. Nonetheless, AFM and optical methods using super-resolution[13] lack direct information about lattice distortions.

Identifying localized structural distortions with x-ray diffraction coupled with x-ray fluorescence spectroscopy is a potential way to investigate the catalytic activity in proximity to defects directly. By combining the analysis with the sophisticated atomic deposition technology of thin films, the impact of inhomogeneities at the hundreds of nanometers length scale can be distinguished from defects such as grain boundaries (often absent in epitaxial films) and vacancy orderings (likely too small to be probed with x-rays). X-ray nanodiffraction enables the study of large distortions in epitaxial thin films, such as phase distribution in materials with a metal-insulator transition[14]. Yet, imaging more subtle localized lattice distortions remains challenging because the signal associated with minute distortions is difficult to isolate and interpret. Here, we combined synchrotron-based x-ray diffraction nanoimaging with machine learning and simulations to categorize heterogeneities in compressively strained $SrIrO_3$ epitaxial thin films, a promising catalyst material for electrochemical conversion in acidic and alkaline solutions[15,16]. We collected 4D scanning data and used k-means clustering to localize regions with lattice imperfections (to 50 nm precision, see SI Methods), which we attribute to strain fields around dislocation half-loops. Furthermore, we find that the strain gradient morphology in a film electrochemically treated under alkaline conditions differs from the morphology in the pristine film. This suggests that electrochemical treatment modifies strain morphology, emphasizing the need for future operando measurements.

Figure 1A shows a schematic of the experimental geometry. The in-plane lattice parameters of the epitaxially grown 12 nm thin $SrIrO_3$ films used in this study are coherently strained to those of the $(LaAlO_3)_{0.3}(SrAl_{0.5}Ta_{0.5}O_3)_{0.7}$ (001) (LSAT) substrate crystal (see diffraction of $103_{pc}$ peak in SI Fig. S1D). Because of the lattice mismatch, the out-of-plane lattice parameter of the film differs from that of the substrate, allowing us to isolate the specular $002_{pc}$ Bragg reflection (pc: pseudocubic), of the film from the much stronger substrate reflection. Figure 1B shows the reciprocal space around the $002_{pc}$ Bragg peak taken with a 0.5 mm x 0.5 mm unfocused x-ray beam in the pristine (as-grown) $SrIrO_3$ film. The well-defined, uniform thickness of the film introduces thickness fringes around the Bragg peak perpendicular to the film surface, and crystal defects – deviations from a periodic crystal – result in diffuse scattering around the peak.

To capture the spatial distribution of the defects, we raster-scanned a focused 30 nm diameter x-ray beam over an area of the $SrIrO_3$ thin film, and recorded 2D diffraction images at each point in the 2D plane of the scan, collectively forming a 4D dataset (Fig. 1A). The scattering geometry used here increases the horizontal footprint of the beam by a factor of $1/\sin(\theta)$, where $\theta = 15.87°$ is the incident angle. We collected scans at three different θ-2θ values to record the diffuse scattering profile in reciprocal space: at the Bragg condition and incident angles offset by $\Delta\theta = \pm 0.2°$ (rocking width ~0.4°). Through varying the incident angle, a portion of broad diffuse scattering is measurable at the same scattering angle $2\theta$, while the position of the sharp Bragg peak changes (Fig. 1C) because the Ewald sphere intercepts the sharp Bragg truncation rod at a different scattering vector Q (Fig. 1B). The highest intensity measured displays a "donut-shaped" ring: a real image of the Fresnel zone plate focusing the x-ray beam (a beamstop blocks the central intensities, and the focusing generates a divergent x-ray beam at the sample[17], and the high quality thin film acts as a mirror in Bragg reflection). At the exact Bragg condition, a complete ring is visible due to the small thickness of the film, resulting in diffraction within a range of incident angles (Fig. 1C, middle). Thus, the intensity contained within the ring primarily comes from the Bragg diffraction condition. In the off-Bragg data, the donut shape exhibits a wide vertical shadow at lower and higher 2θ (Fig. 1C and 1D, left and right, respectively) due to the minimum between the Laue oscillations present in diffraction from a high-quality thin film (Fig. 1B). The diffuse scattering is more discernable in the off-Bragg measurements as a broad background outside the ring, with total intensity two orders of magnitude lower than the donut-shaped peak.

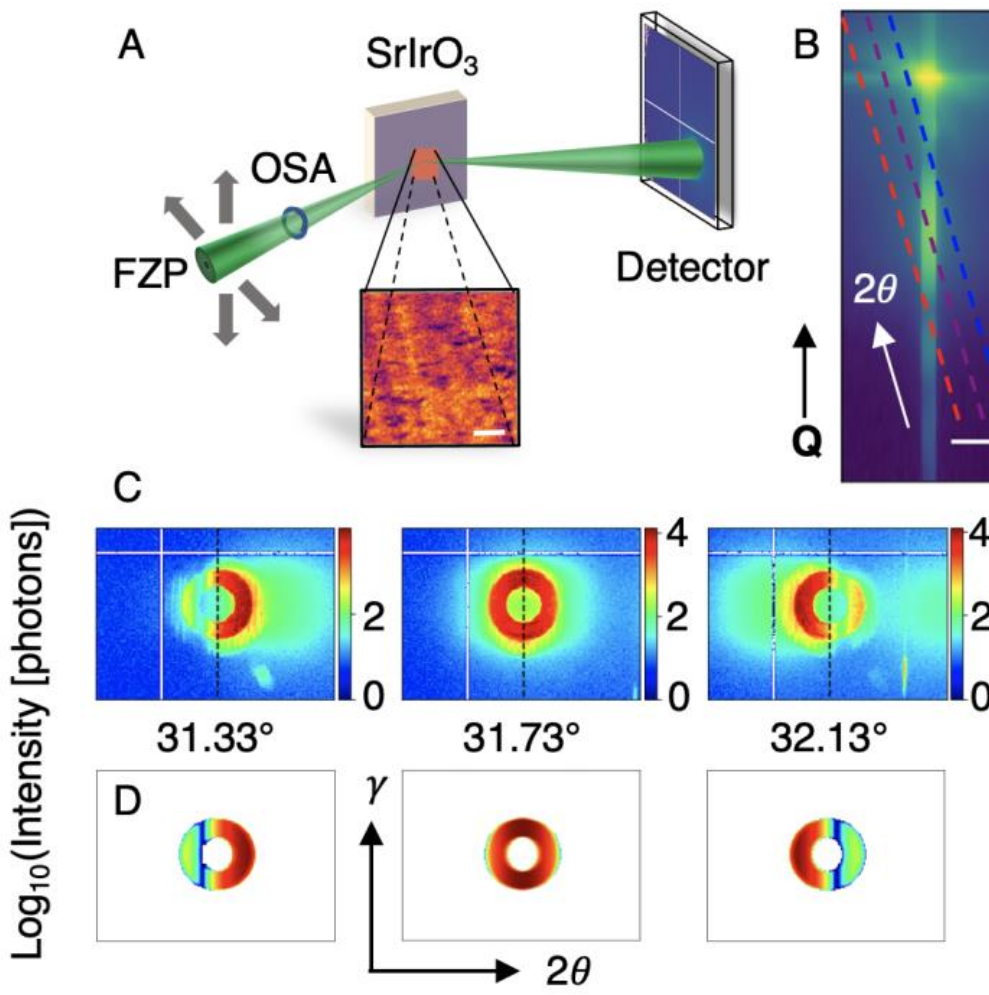

***Figure 1***. *A) Schematic illustration of the focused nano-probe diffraction geometry (scale bar: 1 μm). B)* $SrIrO_3$ $002_{pc}$ *Bragg peak with lines indicating the different cuts of the Ewald sphere as measured at incident angles of 15.67° (red), 15.87° (purple), and 16.07° (blue). Scale bar: 0.11* $nm^{-1}$. *C) Total diffraction intensity integrated over a raster scan across a region of the sample at incident angles of* $\theta$*=15.67° (left),* $\theta$*=15.87° (middle), and* $\theta$*=16.07° (right). Each diffraction image spans 1.32 degrees in* $2\theta$. *D) Simulated total diffraction intensity from a perfect film normalized to the experimental data, where the vertical drop in intensity (left, right, visible as a blue stripe in the false color images) shows the first minimum in the Laue oscillations originating from film thickness. The diffuse scattering visible in the experimental data is missing in the simulation.*

One of the primary benefits of the increased brilliance of synchrotron light sources and upgraded photon detectors is the high speed at which data are collected. The massive amounts of data present a challenge in x-ray science: efficient and effective data processing. Figure 1A shows the map of a 2D area of the pristine $SrIrO_3$ thin film obtained by summing the total intensity from the 2D diffraction pattern collected at each spatial position, thereby creating a four-dimensional (4D) dataset (2D detector image at every step of a 2D raster scan) from single angle diffraction. While the integrated intensity displays some features in the film, it lacks the signatures of the features in reciprocal space. This method of condensing 4D data into 2D loses the nuance of the full diffraction patterns. The scattering signal is inhomogeneously distributed across the measured slice of the Ewald sphere within one diffraction pattern, let alone across the extended spatial region of the sample. Often, nanodiffraction data analysis resembles dark-field imaging with manual identifying regions in the reciprocal space corresponding to different local structures[14]. Yet, the method is inapplicable to the low-signal diffuse scattering.

Here, we used k-means clustering – an unsupervised machine learning algorithm – to categorize the data by the intensity at each pixel position on the detector. K-means clustering is a converging vector quantization algorithm that sorts observations into an integer number, *k,* of clusters[18], and has been implemented for analysis of 4D STEM data[19] and x-ray nanodiffraction of ferroelectric thin films[20]. Following a standard Python implementation of the algorithm[21] with each pixel on the x-ray detector as an "observation" and each position in an area map as a "feature", we clustered the pixels into groups representing different portions of the scattering signal.

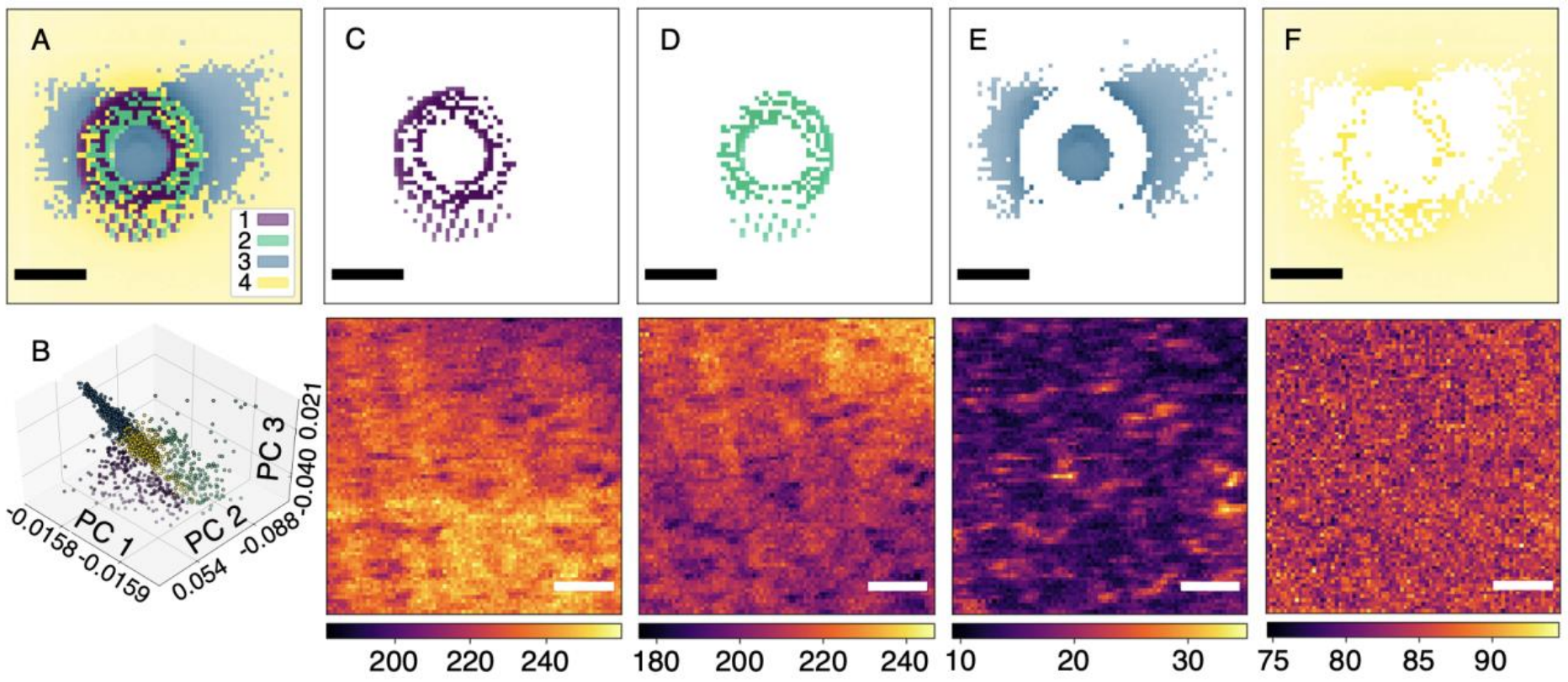

***Figure 2***. *A) Cluster labels for each detector pixel of diffraction at the $SrIrO_3$ $002_{pc}$ Bragg angle ($\theta$=15.87°) using k-means++ centroid initiation. B) Cluster labels plotted along the first three principal components. C-F) Each cluster (top) and corresponding 5 μm x 5 μm spatial map (bottom) generated by the integrated intensity from pointwise multiplication of a binary mask of the cluster and the full diffraction image at each point of the scan. 2$\theta$ is in the horizontal direction, and the scale bar for all diffraction clusters is 0.22°, while the scale bar for all spatial maps is 1 μm.*

Taking the minimum inertia (sum of squared Euclidean distances of observations to their closest cluster center) result of k-means clustering with 9000 randomized centroid initiations, we grouped the diffraction signal into 4 clusters (Fig. 2A), which are orthogonal by nature due to their non-overlapping coordinates in reciprocal space. Figure 2B depicts the clusters in three-dimensional space, with the axes given by the first three principal components of the data. Principal component analysis (PCA) is another example of dimensionality reduction with orthonormal principal components, and can be used to evaluate the quality of clustering[22]. In our case, the four clusters are distinguishable from the PCA condensation of 81 x 81 measurements (dimensions) of 63 x 63 pixels (raw data binned, combining 4 x 4 pixels into 1) into three dimensions. Clusters 1 and 2 (Fig. 2C, D) consist of signals from the donut-shaped zone plate reflection on the Bragg peak, representing scattering from the perfect crystalline lattice. The shift to higher 2$\theta$ (measured as the horizontal position on the area detector) from cluster 1 to cluster 2 indicates a slight tilt or a strain gradient in the lattice planes, indistinguishable using single angle diffraction. The lines below the ring occur due to parasitic illumination from imperfections in the focusing optics. Cluster 3 is diffuse (less structured and broader in reciprocal space), and notably, the first and third principal components (Fig. 2B) shows the separation of Bragg and diffuse scattering, confirming the results of labeling by k-means. Cluster 4 (Fig. 2F) shows no structure, and we attribute it to the background noise. The noise intensity is comparable to the diffuse scattering intensity, demonstrating the algorithm's strength in interpreting noisy data. Furthermore, the background cluster serves as a guideline for the number of clusters used to classify the scattering signal; increasing the number of clusters, *k,* results in the separation of the background signal into different clusters (Fig. S6, S8), a result with no physical grounding.

To relate the different portions of the scattering signal to the positions of nano-scale lattice distortions in the film, we created a binary mask for each cluster (Fig. 2C-F, top), which represents a distinct portion of reciprocal space. The pixel positions given by specified cluster labels were assigned a value of 1, while all other pixels (belonging to different clusters) were set to 0. We then applied this binary mask pointwise to the raw diffraction data, thereby isolating each cluster's signal into a series of corresponding spatial maps (Fig. 2C-F, bottom). This procedure is reminiscent of dark-field imaging except that the unsupervised k-means clustering determines the regions of the masks from the 4D diffraction data. The maps corresponding to clusters 1 and 2 represent regions with high crystallinity and have similar integrated intensities. The slight shift in the reciprocal space indicates a strain gradient or tilt across the top-right corner of the mapped region. The map corresponding to cluster 3 reveals localized lattice distortions as diffuse scattering arises from crystal distortions or defects[23]. Notably, the map of cluster 3 is anti-correlated with both maps from clusters 1 and 2 (perfect crystal): each measured location on the film generates either Bragg or diffuse scattering.

Varying the incident angle allows us to measure the distribution of the diffuse scattering in reciprocal space and its corresponding dark-field maps in real space. At the Bragg condition, the spread of diffuse scattering is centered around the donut-shaped Bragg peak (Fig. 2E). Thus, a significant portion of the measurable diffuse scattering overlaps with the reflection of the Fresnel zone plate. Figure 3 shows the resulting diffuse scattering isolated via k-means clustering (other clusters shown in SI Fig. S5) from scans collected offset from the Bragg peak by $\Delta\theta = \pm 0.2°$ (data in Fig. 1C). The spatial maps found through k-means clustering at different angles are correlated, confirming that the same spatial region of the sample was measured at different angles and that the broad diffuse scattering is visible at off-Bragg angles. Nevertheless, nuanced differences appear: the integrated intensity of the diffuse scattering at the Bragg condition is higher than in both off-Bragg measurements suggesting the diffuse scattering is maximized at the Bragg angle and reduces when scanned perpendicularly to Q. The strength of cluster analysis is further highlighted through the similarity across all three dark-field images in Figure 3: although the diffuse cluster in the on-Bragg scan contains intensity contribution from optics-induced parasitic illumination, it still identifies the same features as in the other two scans with relatively low overlap. Additionally, k-means clustering can be applied to categorize differences within the diffuse scattering at a particular scattering condition (Fig. 3C): the intensity away from the crystal truncation rod distributed perpendicular to the scattering vector indicates in-plane lattice disorder[24,25].

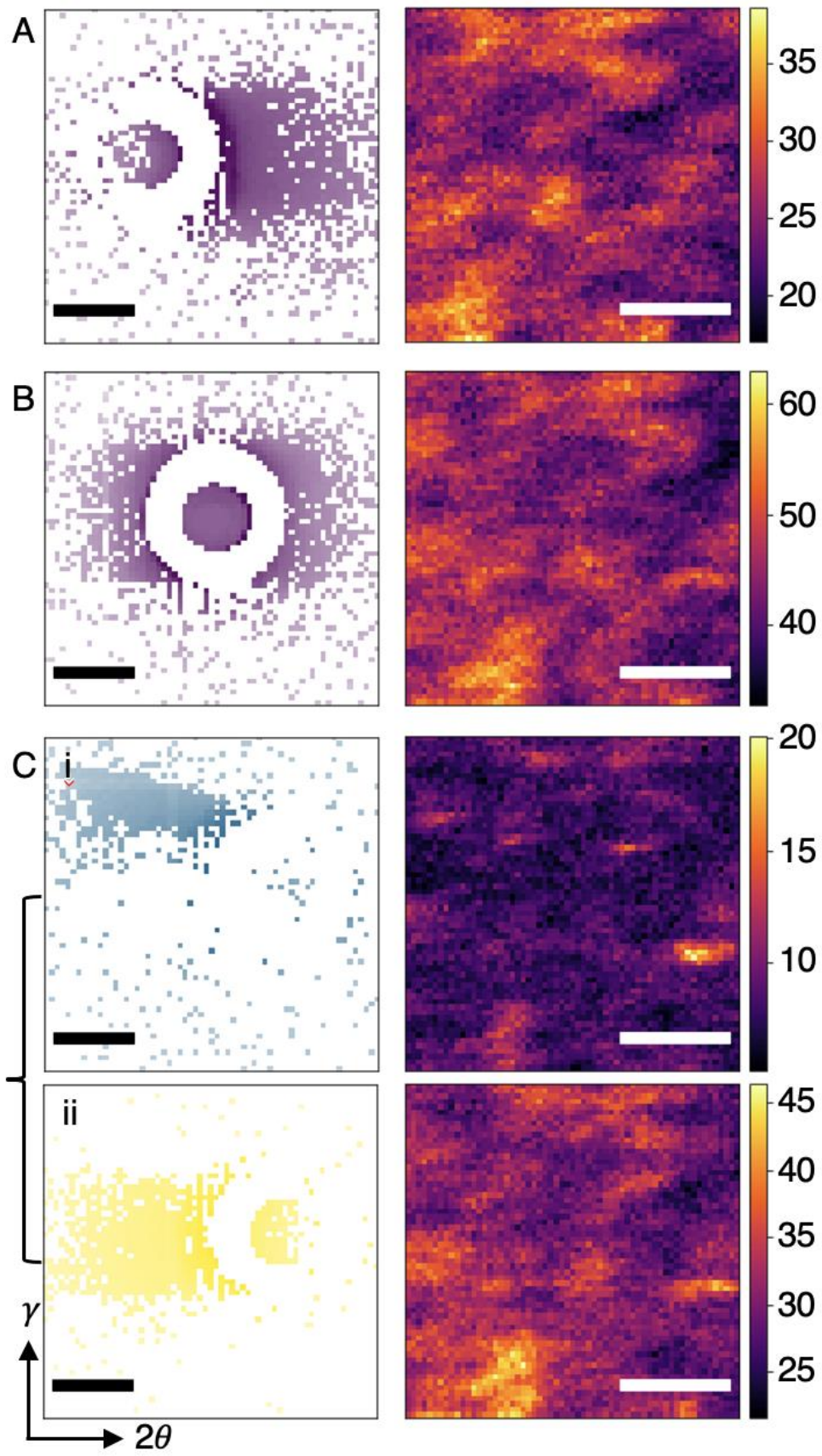

***Figure 3***. *Diffuse scattering identified via k-means clustering (left) of measurements at θ=15.67° (A), θ=15.87° (B), and θ=16.07° (C), and the corresponding integrated intensity maps (right) of the pristine $SrIrO_3$ film. In (i) the diffuse scattering is concentrated in the higher γ, and in (ii) it is spread along the direction of the scattering vector. The scalebar for diffuse clusters (left) is 0.22°, and the scale bar for spatial maps (right) is 1 µm.*

While the current data are insufficient to attribute definitively the observed lattice distortions, we can use their spatial distribution and their signal in reciprocal space to investigate their origin. In epitaxial thin films, local lattice distortions can arise from point defects, yet localized strain due to point defects is likely too small to be imaged by a hard x-ray nanoprobe. Other possible explanations of the observed lattice distortions are threading dislocations and dislocation half-loops, extensively studied in epitaxial semiconductor thin films[26,27] and other functional oxides[28,29]. In heteroepitaxial systems with highly mismatched lattices, misfit dislocations nucleate at the substrate-film interface and propagate normal to the film-air interface. In $SrIrO_3$ grown on LSAT, a system with a relatively low mismatch, half-loops are more likely to nucleate at the film surface and undergo dislocation climb to the substrate-film interface during growth[30]. Measurement of an asymmetric peak with an in-plane component confirms that the film is commensurately strained (Fig. S1C), and thus does not have appreciable density of misfit dislocations. Nonetheless, there is a high concentration (~$10^4$ $cm^{-1}$) of dislocation half loops in epitaxial systems of similar strain[31,32], as well as dislocations and other defects from the substrate crystal, both of which affect local structure and thus scattering intensity in the film.

In the pristine films, the bright regions in Figure 2E (bottom) correspond to areas of relatively high integrated diffuse scattering intensity. These features are several tens to hundreds of nanometers in scale, which correlates well with the expected extension of strain effects due to dislocations. The latter have an average separation of about several hundreds of nm to 1 μm[31,32], accounting for the ability to probe the entire thickness of the film with x-rays. We only measured the $002_{pc}$ specular reflection sensitive to lattice displacement along $[002]_{pc}$. Nonetheless, because edge and mixed dislocations produce strain fields both along the dislocation Burgers vector and perpendicular to it[33], stress induced by such an internal dislocation is relieved into both in-plane and out-of-plane constants locally relative to a fixed Poisson ratio for the unit cell. Thus, our measurements are sensitive to their presence despite the reflection choice. We cannot completely rule out other origins for localized crystal distortions. Nevertheless, the homogeneous distribution of the fluorescence signal (see Fig. S2-S4) indicates the absence of large changes in Ir concentrations or impurities on the surface.

While understanding localized structural heterogeneities in as-grown films provides an important baseline measurement, the greater challenge lies in observing defects in electrochemically cycled $SrIrO_3$ films. During the OER in alkaline electrocatalysis, $SrIrO_3$ adsorbs and transforms surface species from the active $Ir^{4+}$ sites[34], changing the local structure of the film surface and immediate sub-surface layers[35]. Thus, by comparing the heterogeneities between pristine and cycled $SrIrO_3$ films, we can infer the potential morphology rearrangements during the OER electrocatalysis on $SrIrO_3$. Figure 4A shows the six clusters resulting from 3000 initiations of k-means clustering, performed on data collected from an alkaline-treated $SrIrO_3$ film (see SI Methods). Four clusters correspond to signal from the Bragg peak. The map corresponding to cluster 5 highlights the features present in the diffuse scattering. Although the primary contribution to the difference in the scattering intensity concentrates across the stripe running diagonally downward from the upper left corner of the measured region, smaller lines are present at an angle to the larger ones. These features are also shown across the four clusters representing different portions of the Bragg peak, particularly the two wider diagonal stripes of relatively high intensity.

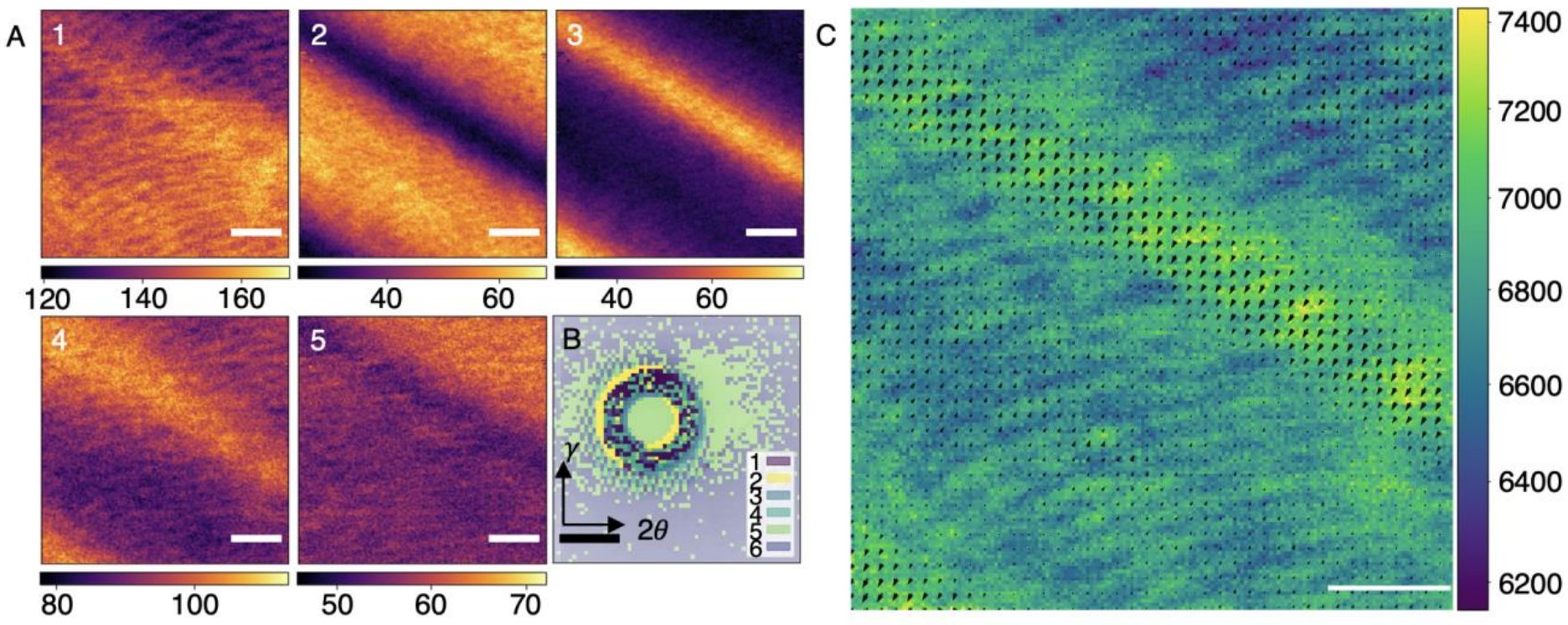

***Figure 4***. *A) 1-5) Corresponding intensity maps to the clusters (B, bottom right). Cluster 6, the background noise, is omitted for brevity. B) Clusters of the nanodiffraction pattern of the electrochemically cycled film collected at the $SrIrO_3$ $002_{pc}$ Bragg condition, with the scale bar indicating 0.22°. C) Integrated intensity map of $SrIrO_3$ $002_{pc}$ of the film cycled electrochemically in 0.1 M KOH, with arrows indicating the simulated direction and relative magnitude of lattice plane tilt. All maps show the same 10 µm x 10 µm measured region (map scalebar = 2 µm).*

Although we cannot directly compare the same region of one sample before and after electrochemical cycling in this study, the measurements before and after differ visibly in morphology. While we identify defects in the as-grown $SrIrO_3$ film as oblong spots 200 to 300 nm in the broad direction, intensity modulations in the alkaline-treated film takes the form of stripes with 600 nm periodicity (Fig. S8). These heterogeneities could be due to a number of factors, including dislocations, amorphization of the surface after electrochemical cycling[35], or uneven film thickness. The featureless Ir fluorescence trace (Fig. S4) confirms that this difference in morphology is due to complex structure within the Bragg peak and extended defects rather than Ir inclusions or cracking of the brittle film. This complex structure highlights the need for analysis methods beyond manual dark-field imaging processing[14], as the subtle differences within the donut-shaped zone plate reflection shown through k-means clustering are impossible to identify and discern manually. The two larger stripes originating from the signal in clusters 2 and 3 (Fig. 4A, 2 & 3, 4B) represent either a tilt or strain gradient, which are indistinguishable in a single Ewald sphere measurement; however, the local strain, tilt, and thickness can be simulated with known sample and experimental geometry parameters (SI, Simulations, Fig. S9)[36]. As shown in Figure 4C the large stripes correspond to regions with tilted lattice planes (~0.01° in the vertical direction and ~0.007° in the horizontal direction), likely due to the miscut of the LSAT substrate crystal. Clusters 2 and 3, as well as the simulated diffraction (Fig. S9) demonstrate how the position of the zone plate image changes in reciprocal space due to lattice rotations. Notably, the intensity corresponding to the Bragg peak (Fig. 4A, 1-4) is not two orders of magnitude higher than that of the diffuse scattering (Fig. 4A, 5) as in the pristine sample (Fig. 2C-E). Within all measured regions of both the as-grown (4 regions of between 3 x 3 and 5 x 5 $\mu m^2$) and electrochemically cycled samples (3 regions), the fine striped structure was only observed throughout the cycled film (SI Fig. S6). Nevertheless, we compared two different sample regions, which highlights a fundamental flaw in *ex situ* measurements – the inability to monitor structural changes in response to an external driving force – and emphasizes the necessity of *operando* measurements.

In this study, we combined unsupervised k-means clustering with scanning x-ray nanodiffraction to identify and characterize defect and strain behavior from the diffuse scattering produced by as-grown and alkaline-treated $SrIrO_3$ epitaxial thin films. We measured localized distortions from the ideal crystalline lattice in pristine films and observed that the defect morphology is different in electrochemically cycled films. Additionally, the presented work demonstrates the potential that unsupervised machine learning applied to the 4D x-ray nanodiffraction datasets has for isolating low-intensity diffuse scattering signal and subtle changes within the Bragg peak. We anticipate that the approach shown will serve as a methodology to monitor localized morphological changes and their role in electrochemical reactions in the future.

**Supplementary Material**

See supplementary material for materials synthesis, electrochemical cycling, x-ray nano-diffraction parameters, and additional results.

**Acknowledgments:**

The authors acknowledge Dr. Ludi Miao for his help in lithography of fiducials, and Matthew Barone for in-house reciprocal space mapping. This work was primarily supported as part of the Center for Alkaline Based Energy Solutions (CABES), an Energy Frontier Research Center funded by the U.S. Department of Energy, Office of Science, Basic Energy Sciences at Cornell under award # DE-SC0019445 (A. L., O. Yu. G., Z.S., D. K., R. B., J.S., A.S.). A.L. acknowledges a graduate research fellowship through the National Science Foundation (DGE-2139899). Use of the Advanced Photon Source and the Center for Nanoscale Materials, both Office of Science User Facilities, was supported by the U.S. Department of Energy, Office of Science, Office of Basic Energy Sciences, under Contract No. DE-AC02-06CH11357. The thin film synthesis was supported by the National Science Foundation (Platform for the Accelerated Realization, Analysis, and Discovery of Interface Materials (PARADIM)) under Cooperative Agreement No. DMR-2039380 (J.N.N, K.M.S., D.G.S.). This work was performed in part at the Cornell NanoScale Facility (CNF), a member of the National Nanotechnology Infrastructure Network, which is supported by the NSF (grant ECCS-0335765).

**Data Availability Statement**

The data that support the findings of this study are available from the corresponding author upon reasonable request.

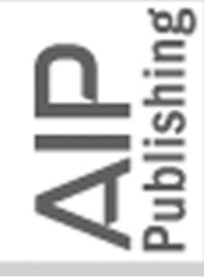

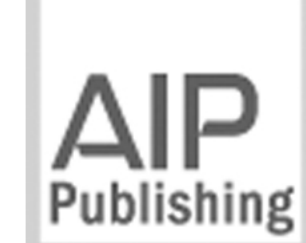

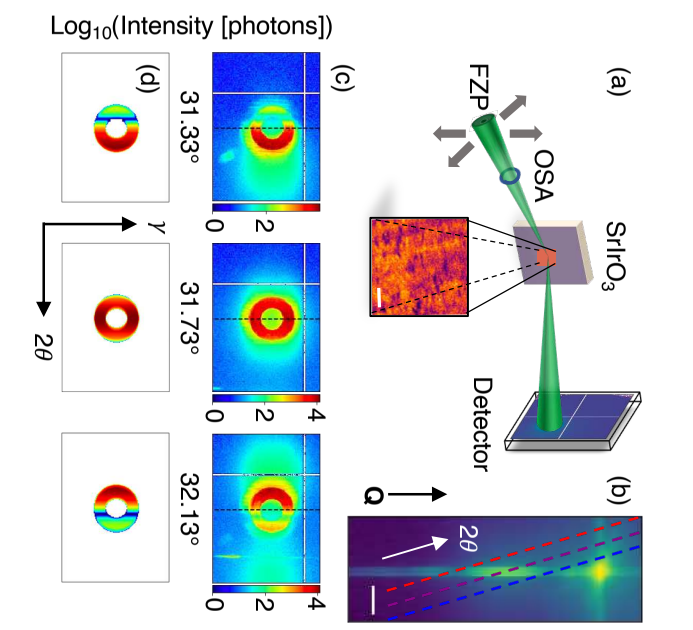

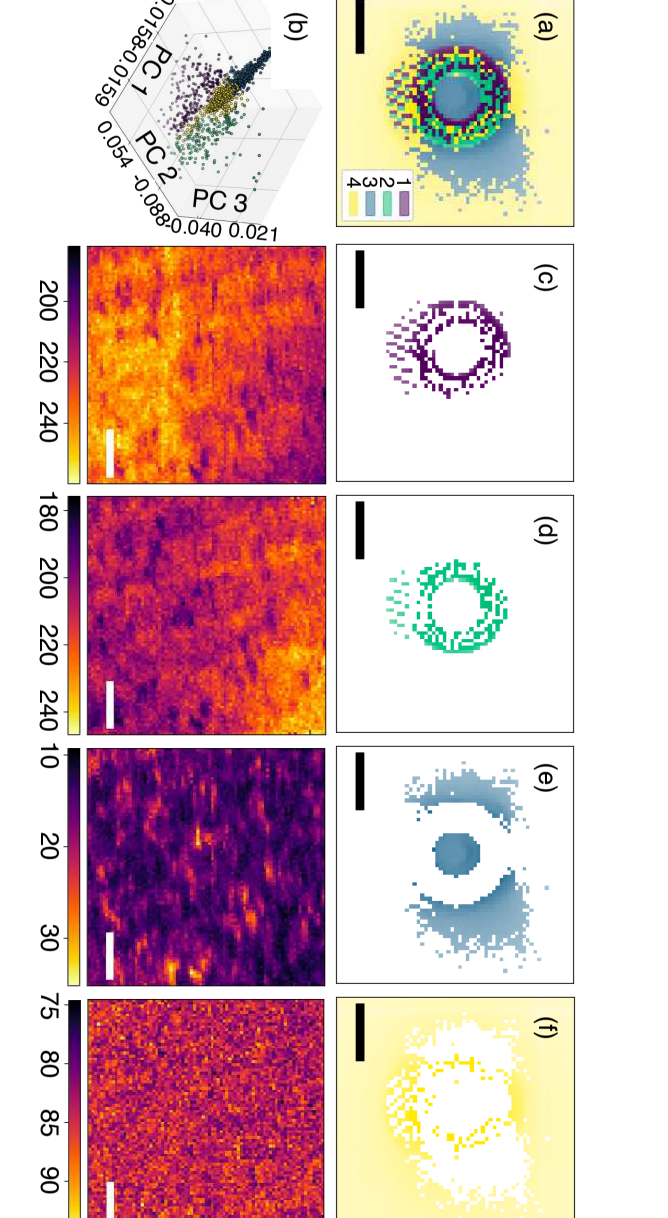

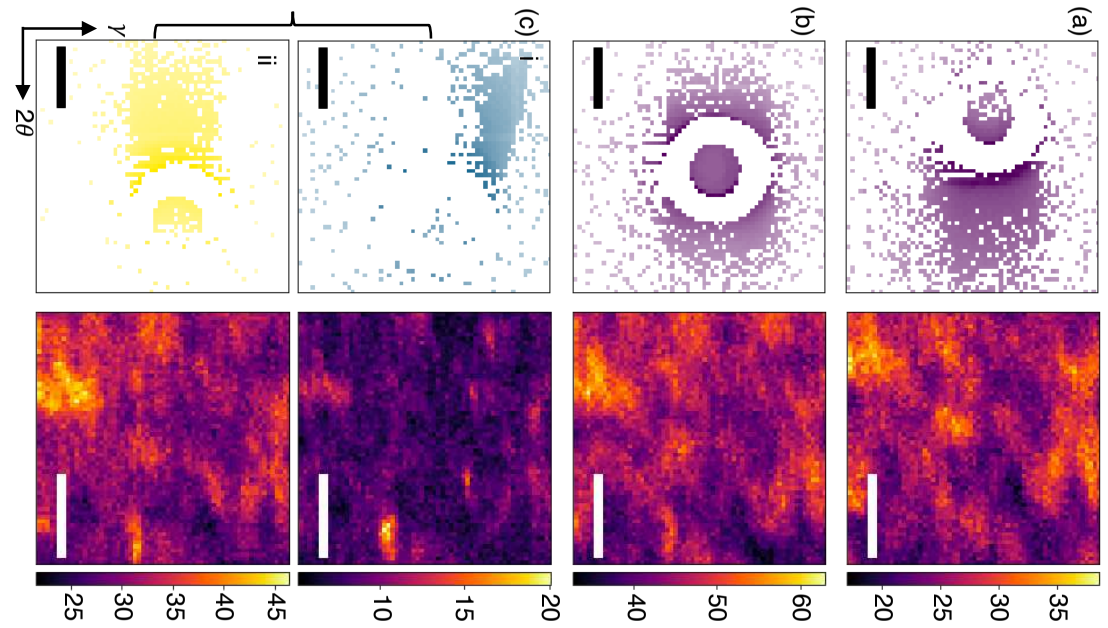

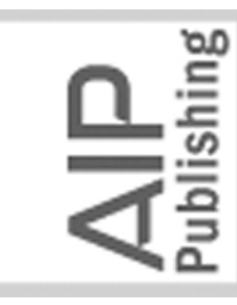

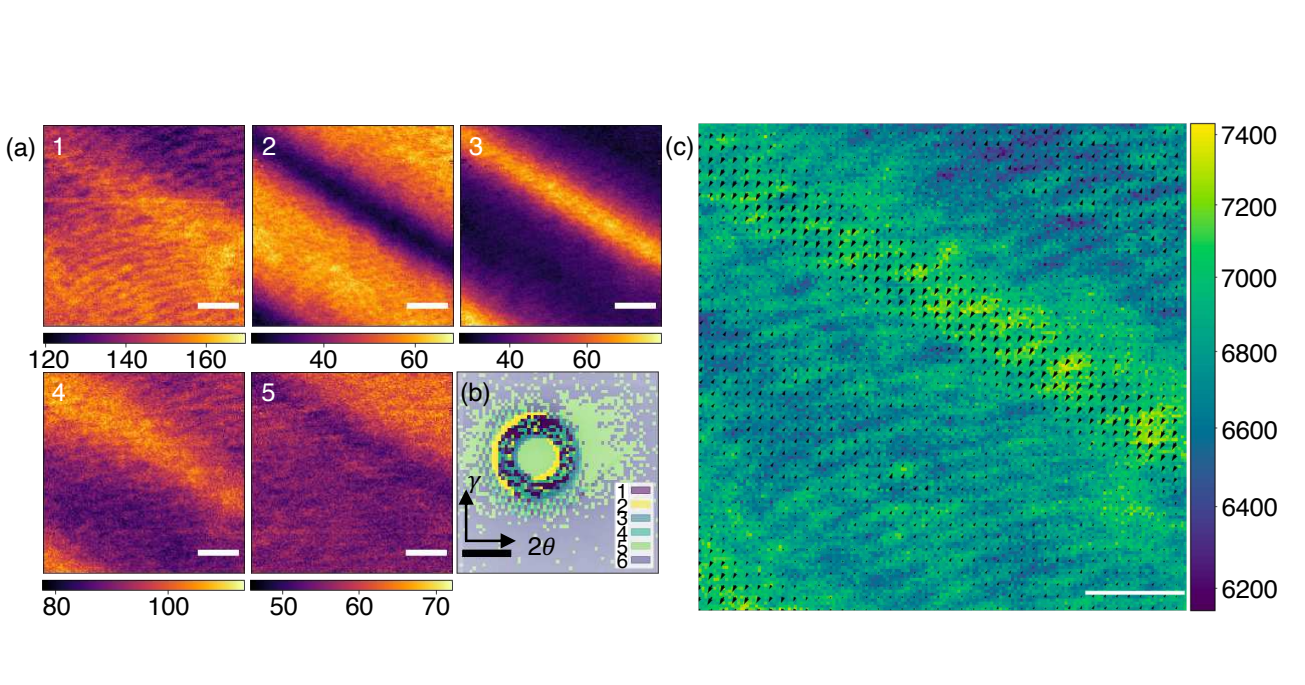